\documentstyle[emulateapj]{article}
\def\gtrsim{\mathrel{\hbox{\rlap{\hbox{\lower4pt\hbox{$\sim$}}}\hbox{$>$}}}}
\let\ga=\gtrsim
\def\lesssim{\mathrel{\hbox{\rlap{\hbox{\lower4pt\hbox{$\sim$}}}\hbox{$<$}}}}
\let\la=\lesssim

\def\chandra{{\it Chandra~}}

\begin{document}

\title{Different Methods of Forming Cold Fronts in Non-Merging Clusters}
\author{Renato Dupke}
\affil{University of Michigan, Ann Arbor}

\author{Raymond E. White III}
\affil{University of Alabama, Tuscaloosa}

\author{Joel N. Bregman}
\affil{University of Michigan, Ann Arbor}

\begin{abstract}

Sharp edges in X-ray surface brightness with continuous gas pressure called cold fronts have been often found in relaxed
galaxy clusters such as Abell 496. Models that explain cold fronts as surviving cores of head-on subcluster mergers do 
not work well for these clusters and competing models involving gas sloshing have been recently proposed. Here, we
test some concrete predictions of these models in a combined analysis of density, temperature, metal abundances and 
abundance ratios in a deep \chandra exposure of Abell 496.   
We confirm that the chemical discontinuities found in this cluster are not consistent with a core merger remnant scenario. 
However, we find chemical gradients 
across a spiral ``arm'' discovered at 73 kpc north of the cluster center and coincident with the 
sharp edge of the main cold front in the cluster. Despite the overall SN Ia iron mass fraction dominance
found within the cooling radius of this cluster, the metal enrichment along the arm, determined from silicon and iron 
abundances, is consistent with
a lower SN Ia iron mass fraction (51\%$\pm$14\%) than that measured in the surrounding regions (85\%$\pm$14\%). 
The ``arm'' is also significantly colder than the surroundings by 0.5--1.6 keV. 
The arm extends from a boxy colder region surrounding the center of the cluster, where two other cold fronts
are found.
This cold arm is a prediction of current high resolution numerical simulations as a result of an
off-center encounter with a less massive pure dark matter halo and we suggest that the cold fronts in A496 provide the first
clear corroboration of such model,
where the closest encounter happened $\sim$ 0.5 Gyr ago. We 
also argue for a possible candidate dark matter halo responsible for the cold fronts in the outskirts of 
A496.

\end{abstract}


\section{Introduction}

One of the most interesting features discovered by {\sl Chandra} satellite observations 
of galaxy clusters are the sharp X-ray surface brightness discontinuities, accompanied by jumps 
in gas temperature named ``cold fronts'' (e.g. Markevitch et al. 2000; Vikhlinin et al. 2001; 
Mazzotta et al. 2001). The temperature and density jumps happen
in such a way as to maintain the gas pressure continuously across the front and, therefore, they are not 
created by shocks. They were  originally interpreted as being the result of  
subsonic (transonic) motions of head-on merging substructures with 
suppressed thermal conduction (Markevitch et al. 2000, 2001; Vikhlinin et al. 2001).

The above mentioned merger core remnant model is 
theoretically justified (e.g. Bialek, Evrard \& Mohr 2002; Nagai \& Kravtsov 2003; Heinz et al. 2003; 
Mathis, et al. 2005; Poole et al. 2006) and holds relatively 
well for clusters that have clear signs of merging, such as 1E0657-56 (Markevitch et al. 2002) 
and A3667 (Vikhlinin et al. 2001). However, these models do not work well 
for the increasing number of cold fronts (sometimes multiple cold fronts in the same cluster) found 
in apparently non-merging clusters such as A496 (Dupke \& White 2003, hereafter DW03), A1795 
(Markevitch et al. 2001), RXJ1720.1+2638 (Mazzotta et al. 2001). This 
 prompted the development of other models for cold front generation,
 such as oscillation of the cD and the low entropy gas around 
the bottom of the potential well (Lufkin et al. 1995; Fabian et al. 2001; DW03), 
hydrodynamic gas sloshing (Ascasibar \& Markevitch 2006, hereafter AM06), or dark matter peak 
oscillation due to scattering of a smaller dark matter system (Tittley \& Henriksen 2005). For 
very recent review see Markevitch \& Vikhlinin (2007).

Cold fronts are found with relatively high frequency. A review of \chandra archival
images finds that more than 40\% of the observed clusters have
cold front-like features and their presence may have 
significant physical impact in the physics of their host cluster cores, such
as gas heating, generation of bulk and turbulent velocities, constraining conduction, etc. 
The significance of cold fronts influence on cluster physics depends on 
how they are being generated. Therefore, it is important 
to determine which mechanisms actually produce cold fronts. Abell 496 provides 
an excellent opportunity to test different scenarios for 
cold front generation given its physical and observational characteristics.
A496 is a typical, bright, nearby (z$\approx$0.032), apparently relaxed cold  
core cluster. The X-ray peak coincides very well with the cD optical centroid.
The gas temperature varies from 5--6 keV in the outer regions to 2--3 keV in the central 
arcmin (e.g. Tamura et al. 2001, DW03).
The presence of a central abundance enhancement has been established with previous instruments including 
{\sl Ginga} and  {\sl Einstein} (White et al.\ 1994), {\sl ASCA} (e.g. Dupke \& White 2000a), {\sl BeppoSAX} 
(Irwin \& Bregman 2001) and {\sl XMM} (Tamura et al 2001), showing an overall radial enhancement from $\sim$0.2-0.25 solar
in the outer regions to $\sim$ 0.4-0.7 solar in the central arcmin. Furthermore, Dupke \& White (2000a) also 
discovered 
radial gradients, for the first time, in various elemental abundance $ratios$, which indicates that the gas in the central 2-3$^\prime$ has 
a higher proportion of SN Ia ejecta ($\sim$70\%) than the outer parts of the cluster. This was confirmed by more 
sensitive spectrometers on-board {\sl XMM} (Tamura et al 2001).

As pointed out by DW03, different models for cold front formation can be discriminated through 
the analysis of chemical gradients across the front. If the cold front is a due to a 
head-on merger core remnant, we should expect the front to be accompanied by a specific  
discontinuity of elemental abundance ratios (e.g. Mushotzky et al. 1996; Dupke \& White 2000a,b). 
The expected discontinuity in this case would be symmetric with respect to the merger axis and asymmetric
with respect to the perpendicular direction to the merger axis.
This kind of analysis can be 
performed best with {\sl Chandra}, given its high angular resolution. 
DW03 performed a chemical analysis of the cold front in Abell 496. With an 
effective exposure of $\sim$9 ksec, 
they were able to determine abundance ratio profiles only on large semi-annuli, covering a region larger than that of 
the cold front itself. The distribution of iron, silicon and oxygen abundances 
showed radial gradients but there were no clear discontinuities uniquely related to the cold front itself, pointing out
the weaknesses of the remnant merging core model when applied to A496. Here we 
report the results of 
a deeper observation of that cluster that allowed us to produce 
high quality maps of the gas parameters 
and to compare more closely the observations with the predictions given by different models for cold 
front formation. All distances shown in this {\it Letter}
are calculated assuming a H$_0=70$ km~s$^{-1}$Mpc$^{-1}$ and 
$\Omega_0=1$ unless stated otherwise. At the distance of this 
cluster $1^{\prime\prime}\approx$ 0.66 kpc. 

\section{Data Reduction}
Abell 496 was observed by {\sl Chandra} ACIS-S3 in July 2004 for 76 ksec.
The cluster was centered on the S3 chip. We used Ciao 3.2.0 with CALDB 3.0 to screen the data.
After correcting for a short flare-like period the resulting exposure time in our analysis
was 59.6 ksec. A gain map correction was applied together with PHA and pixel randomization. 
ACIS particle background was cleaned as prescribed for VFAINT  
mode. Point 
sources were extracted and the background used
in spectral fits was
generated from blank-sky observations using
the {\tt acis\_bkgrnd\_lookup} script. 
Here we show the results of spectral fittings with XSPEC V11.3.1 (Arnaud 1996) using the 
{\tt apec} and {\tt Vapec}
thermal emission models. 
Metal abundances are measured relative to the solar photospheric values of 
Anders \& Grevesse (1989).
Galactic photoelectric absorption was incorporated using the {\tt wabs} 
model (Morrison \& McCammon  1983).
Spectral channels were grouped to have at least 20 counts/channel. Energy 
ranges were restricted to 0.5--9.5 keV. The spectral fitting errors 
are 1-$\sigma$ confidence unless stated otherwise.

In order to obtain an overall distribution of the spectral parameters we used an 
adaptive smoothing code that selects extraction regions based on a fixed minimum number of counts
 per cell (here we used 3000 counts for temperatures and global abundances 
 and 7000 for individual abundances) to maintain the range of statistical fitting errors more or less 
 constant throughout. The intercell spacing is fixed at a fraction of the radius of the surrounding cells
 and in general there is significant 
 cell to cell overlap except for the cells with smallest size. 
 The overlap of extraction regions is therefore stronger in 
low surface brightness regions, away from the core of the cluster. We plot the distribution of 
region sizes in Figures 3e and 4e, to give an estimate of the local smoothing kernel size.
The code produces a matrix with best-fit values and different cell sizes. The best-fit values
used here are defined as the mid point of the 68\% confidence errors. 
In order to make the contour plots, this matrix is mapped into a square matrix with equal cell sizes 
using an interpolation routine. This is done by computing a new value for each cell in
the regular matrix weighing by the values of the adjoining cells in the matrix included within 
some defined search radius (minimum of 3 cells in 4 adjacent quadrants). The closest measured 
values usually have the most influence on calculating the value of a cell. 
The computation is based on the Kriging method (for a description see, e. g., Davis, 1986, p.383),
which calculates the weights from a semivariogram ($\gamma(h)~=~\frac{\sum_{i}( X_i~-~X_{i+h})^2 }{2n}$) 
developed from the spatial structure of the data, where $h$ is the number of intervals between the values
of the regionalized variable $X$ taken at location $i$ and $i+h$ and $n$ is te total number of points.
The number of cells of the mapped matrix was artificially increased to three times of the maximum 
length of the original matrix for purposes of improving image quality for analysis. This is responsible 
for the small ``square domains'' that appear in Figures 3a,c \& 4a,c. The values outside the CCD
border contours are also an effect of the smoothing algorithm and should be ignored.

\section{Results}
\subsection{Cold Fronts and Temperature Distribution}
Figure 1a shows the exposure corrected smoothed X-ray image of A496. 
One can clearly see the sharp surface
brightness edge towards the north, described in DW03. One can also see two other brightness 
edges (to the SW and SE) that meet at nearly right angles. This suggests the presence of multiple 
cold fronts in this cluster\footnote{Tanaka et al. (2006) finds an additional cold front in this cluster 
4$^\prime$ to the south, out of the field of view of our observation}.
To analyze the nature of these edges we used the set of extraction regions shown in Figure 1b.
The results are shown in Figures 2a,b using a {\tt wabs apec} spectral model. 
Figure 2a shows the distribution of surface brightness (top) and
projected gas temperatures using the bins shown in Figure 1b (bottom) . The color association between the 
Figures 2 and Figure 1b is: North--black, East--red, South--blue and West--green. The locations of the cold
fronts are marked by vertical dashed lines and follow the same color code. 
There are at least three (up to five)
surface brightness edges accompanied 
by sudden temperature jumps, consistent with cold fronts; The northern one  is at 
$\sim$ 73 kpc and is the strongest. The western cold front is nearly at the same radial distance 
(r $\sim$ 64 kpc) as the northern one and is,  
apparently, an extension of the northern front. We can also see that the two edges
near the core, labeled East (r $\sim$ 16 kpc) and South (r $\sim$ 22 kpc)
have the temperature jumps
characteristic of cold fronts. There is also a
marginally significant cold front  to the east at r $\sim$ 106 kpc. 

Following DW03, we measured the 
radial distribution of metal abundance ratios towards the directions of the main fronts (edges). 
We used ``PIE'' extraction regions that were chosen in such a way as to have the same opening angle
as the cold front of interest.
In the radial distributions there are no clear significant 
systematic relations 
between the changes in Fe abundance or abundance ratios and cold fronts. 
The changes seen can be mostly 
associated with overall (global) radial trends. 
Globally, the Fe abundance shows a radial decline 
from supersolar near the cluster's center to subsolar in the outer core regions. 
At the very center, r$\la$10$^{\prime\prime}$ ($\sim$~7 kpc), there is a
significant abundance dip described in the next section. 
The radially average values in the central 23 kpc is  0.93$\pm$0.04 solar (with asymmetric 
variations from 0.8 solar to 1.2 solar)
and in the outer (130$\pm$50) kpc is 0.75$\pm$0.04 (with asymmetric variations from 0.47 solar to 0.86 solar).


The results for the ratios involving Si,
S, and Fe are shown in Figure 2b, where the color code is the same as that used in Figure 2a.
In the abundance ratio plots we added ``2'' to the values of the Northern and Western directions, 
for illustration purposes.
Despite the strong anisotropies, particularly in the bins within $\sim$70--110 kpc, 
there is also a tendency for the $\alpha$--element ratios to Fe grow radially.
The Si/Fe abundance ratio is consistent with a flat (or mildly increasing) profiles going
from 1.39$\pm$0.10  in the central $\sim$11 kpc to 
1.52$\pm$0.23 in the very outer core regions ($\ga$ 110 kpc). 
In the same regions, the S/Fe ratio exhibits a more significant gradient, 
growing from $\sim$ 1.46$\pm$0.14 solar 
 to 2.34$\pm$0.38 solar. 
The error weighted average of the SN Type dominance of the Fe mass from the two ratios above corresponds
to 65\%$\pm$4\% SN Ia Fe mass fraction for the central region and 57\%$\pm$7\% in the outer regions. This is
consistent with the general trend found with {\sl ASCA} by Dupke \& White (2001a) for larger spatial scales (up to 
$\la$ 500 kpc, although the absolute values of the sulfur abundance are higher than those
determined with {\sl ASCA} (Dupke \& White 2000a) and {\sl XMM} (Tamura et al. 2001)\footnote{It should be noted 
that even though the radial trends found for ratios that include S are, in general, similar to those derived
from other ratios, the absolute  
values of the S abundance seem to be overestimated with respect to SN Ia and II yield models. 
To place it within the theoretical models 
the values measured would 
need a systematic negative correction. This does not change the conclusions of this paper since we are 
looking at {\it relative chemical} 
changes across cold fronts. For a discussion about the discrepancies found between observed 
sulfur yields and model predictions see 
,e.g., Dupke \& White 2000a; Baumgartner et al. 2005 and references therein)}.

The variations in the radial distributions of abundance ratios and temperatures suggest the presence
of significant asymmetries. To explore the 
nature of these asymmetries we produced 2--dimensional adaptively smoothed maps of projected gas 
temperatures, abundances and abundance ratios. We discuss them in the next section.

\subsection{2-D Maps}

Given the level of asymmetry of the distributions of gas temperature, metal abundances and abundance ratios,
it is helpful to analyze the 2-D distributions of these parameters. The temperature and Fe abundance maps are
shown in Figures 3a,c, with X-ray surface brightness contours used in Figure 1b overlaid.
The steepest temperature gradient is seen to the North. One striking feature 
that can be seen in the temperature map is a ``cold spiral arm'' that departs from the core to the N-NW up to the cold 
front position and runs along the cold front to the E-NE becoming more diffuse as it turns towards the S. 
The smoothing kernel radius map (Figure 3e) shows a value of $\sim$20(30) pix or 10$^{\prime\prime}$(15$^{\prime\prime}$)
in the inner (outer) arm regions, which is nearly half of the arm thickness and indicates that the arm is well spatially resolved. 
Guided by the temperature map, we defined regions that characterize the inner and outer parts of the arm for 
spectral extraction and they are shown in Figures 3a,c \& 4a,c, and the relevant best fit parameters are shown in Table 1.
The temperature of the cold arm is $\sim$3.08$\pm$0.07 keV.
The temperatures on the surrounding regions of the cold arm are 3.5$\pm$0.11 keV and 4.7$\pm$0.22 keV towards the inner and outer 
cluster regions, respectively.
The cold arm is definitely associated with the northern cold front and to a lesser extent to the 
western cold front. It departs from a boxy low temperature region, the edges of which 
appear coincide with the southeastern and southern cold fronts near the cluster's core, 
although the temperature edges in these weaker cold fronts are less well-defined 
than that of the main cold front. From Figure 3b it can be 
seen that the overall temperature error
in the cold arm region is around 0.1--0.2 keV. The higher 
temperatures near the southern CCD border are not well constrained (with errors $\ga$ 1 keV).  

There are significant indications of a ``cold tail'' (T $\sim$ 4 keV ) starting 
2$^\prime$.3 southwest of the cluster's center extending to 4$^\prime$.2 to the 
south of the cluster that is associated 
with a low Fe abundance region (Figure 3c). The abundance along the cold tail is 
approximately half of the surrounding regions values of $\sim$1.2 solar. 
This ``cold tail'' seems to extend to the south for more 5$^\prime$ 
(Tanaka et al. 2006).
A similar 
cold tail was found on the opposite side of the cold front in the cluster 2A0335+096 (Tanaka et al. 2006). 

The Fe abundance map is also inhomogeneous (Figures 3c, d). There is an overall abundance gradient,
which is steeper towards the northern regions. In particular the transition from sub to super
solar abundances happens at a radius of 100$^{\prime\prime}$--140$^{\prime\prime}$ from the center 
in all directions but the South. In general, the Fe abundance within 
the main cold front spatial scales (r$<$60 kpc) is supersolar, with the exception of the 
very central 8 kpc, where an abundance ``dip'' is found. The Fe abundance in the central 
dip reaches a minimum of
0.55$\pm$0.3 solar (an average 0.8$\pm$0.03 solar in a circular region 10.5 kpc in radius)
and in the immediately surrounding regions achieves a maximum of $\sim$1.7$\pm$0.4 solar 
(an average of 1.1$\pm$0.04 solar within an annulus with radius between 
11 kpc and 22 kpc). There is a secondary, marginally significant, abundance 
dip with similar spatial scales 
35$^{\prime\prime}$ to the N-NW, where the abundance decreases from $\sim$1.3 to $\sim$0.7 solar 
with a characteristic error of 0.3 solar. 
Central metal abundance dips have been found in other clusters (e.g., A2199 (Johnstone et al. 2002), 
Centaurus (Sanders \& Fabian 2002) and Perseus (Schmidt et al. 2002)),
and the mechanisms that generate them are a matter of current debate. Suggested scenarios 
include resonant scattering (cf. Sanders \& Fabian 2006), extremely inhomogeneous
metal abundances (Morris \& Fabian 2003), artifacts appearing from 
fitting single temperature models to multi temperature gas (Buote 2000) and buoyant transport to higher radii 
(Brighenti \& Mathews 2005). None of these mechanisms are adequate to explain off-center abundance
dips, which are probably related to previous AGN activity. A extended analysis of off-center abundance 
dips in clusters is provided elsewhere (Dupke, Nyland \& Bregman 2007, in preparation).
Metal abundances are in general high 
towards the southern regions, with the exception of the regions coincident with the southern cold tail.

We performed an analysis of the 2-D distribution of the elemental abundance ratios in this cluster. Different 
metal enrichment mechanisms act with different efficiencies at different cluster locations and produce different 
SN type ejecta signatures. Therefore, elemental abundance {\it ratios} can be used as ``fingerprints'' used to trace the gas 
history, better than metal abundances alone. The abundance ratio maps involving the best determined abundances 
(Si, S, and Fe) are shown in Figures 4a, c. The 1-$\sigma$ errors of the quantities are shown in Figures 4b, d, 
and give an idea of the significance level of the measured quantity in the region of interest. 
Since our best-fit values are defined as the mid-point of the 1-$\sigma$ error bars, we use
only values with  fractional errors smaller than 100\% were used to create the 2-D square images.
This is done to avoid biases in the interpolation to produce the smoothed color contours that 
would be caused by upper/lower limits, where the error bars can be highly asymmetrical. 
It can be seen that, in general, the cold arm is accompanied by 
enhanced abundance ratio
values (lower SN Ia Fe mass fraction than the surroundings), which is visible in the Si/Fe, which 
shows an average variation 
from $\sim$1 to 2, or equivalently, from 
85\% to 51\% SN Ia Fe Mass fraction, respectively in the regions surrounding the cold arm and 
the regions along the cold arm. The characteristic error is $\sim$0.4 ($\sim$14\% in SN Ia Fe mass fraction) and the characteristic 
smoothing kernel size is $\sim$30 (50) pixels, or 15$^{\prime\prime}$(25$^{\prime\prime}$). Sulfur abundances are higher 
than expected and abundance ratios are
off-scale when compared to the theoretical predictions of Nomoto et al. 1997a,b for SN Ia and II yields.
However, the trend of S/Fe is similar to that of Si/Fe and to place the  limits within 
theoretical bounds, we need to apply constant positive correction of $\sim$0.4 to S/Fe, 
placing the  and the corresponding S negative correction $\sim$ 0.4--0.8 within the errors (see footnote 2).

\section{Discussion: The Nature of Cold Fronts in Abell 496}

The analysis of the core of A496 presented in this {\it Letter} reveals several new features that
were not observed previously. A large multiplicity of cold-front features (at least three 
cold fronts); a spiral cold arm seen in the temperature map, which is clearly associated with the main (northern) cold front; 
strong indication of spiral (or circular) chemical arms associated with the main cold front; 
a cold, metal poor tail extending towards the direction opposite to the main cold front; an overall
central abundance enhancement with a small-scale ``dip'' at the core, and marginal evidence for 
other off-center abundance dips. The multiplicity of cold fronts together 
with the spiral pattern of the chemical gradients seem to rule out the scenario, where the 
cold front(s) in this cluster are created by a head-on merging remnant core. 

Although gas sloshing has been invoked to explain cold fronts in apparently relaxed clusters,
there have been very few observable predictions that can be used to discriminate the details of
different sloshing mechanisms proposed in the literature. 
Very recently, AM06 performed 
high resolution numerical + hydrodynamical simulations specifically designed to investigate the effects 
of scattering of lower mass dark matter haloes (with and without gas) by clusters of galaxies.
One of the results from their work was that the sub-halo flyby induces a variable gas velocity field in the
ICM of the main cluster that generates ram-pressure near the cluster gas core and produces cold fronts, 
accompanied by significant amount of substructures seen in the gas 2-D temperature distribution.

 A common feature in most cases analyzed by AM06 was the 
presence of cold spiral arms coinciding with the cold fronts close to the main cluster's core, which were long lasting.
In particular, their case for
a dark matter perturber produces properties very similar to those observed in A496. 
In AM06 a pure dark matter halo with $\frac{1}{5}$ of the mass of the main cluster flies by with an impact parameter 
of 500 kpc and with closest approach at t$\sim$1.37 Gyr. We show part of Figure 7 of AM06,
for the epoch corresponding to 1.9 Gyr (Figure 5a).  The image is inverted vertically to be 
compared directly to the temperature map 
of A496 in Figure 2a. The size of the box is 250 kpc, similar to the size of ACIS-S3 CCD borders at the redshift
of the cluster ($\sim$320 kpc). The cold front(s) can be seen when comparing the temperature map with the surface
brightness map (Figure 21 of AM06). The main cold front coincides with the large spiral cold arm
extending horizontally. The spatial scale is very similar to that of the cold
arm in A496. Their simulations also seem to indicate the presence of milder cold fronts in the opposite
side closer to cluster's core. These are clear predictions that are corroborated well by A496 and suggest 
strongly  that 
a flyby dark matter halo created the cold fronts in this cluster. 
Furthermore, there is a larger-scale more diffuse cold extension of the main arm also towards the 
South of the main cold front, which is a consequence of the 
ram-pressure caused by the gas velocity field induced by the DM halo flyby. This suggests that the 
same process that creates the main cold front may also be associated with the formation of the 
southern cold tail seen in A496. 
The existence of such pure DM sub-halos is not 
completely unexpected since the intergalactic gas originally belonging to the sub-halo could have been
stripped in a previous encounter with the main cluster. AM06 cases for {\it gaseous} DM sub-clump passages
produces a variety of substructures visible in temperatures {\it and surface brightness} maps, which are not 
seen in A496, and are not favored within the limited cases simulated. Future addition of metallicity 
distributions to cluster merger simulations should 
constrain further the characteristics of the perturber.

A prediction of this scenario is the presence of a DM halo in the outskirts 
of the cluster without significant X-ray emitting gas. From the simulations, 
the position of that clump at epoch (t=2 Gyr, i.e., now) 
would be towards its apocenter at North, the same general direction of the main cold front. 
It is reasonable to assume that galaxies would tend to trace their host DM sub-halo.
 Recent wavelet analysis of the member galaxies of A496 within a 1.5 h$_{75}^{-1}$ Mpc radius 
(Flin \& Krywult 2006),
finds a secondary galaxy clump, in most wavelet scales 
analyzed, to the NW of the core of A496, roughly consistent with the position where 
the DM perturber was likely to be found in the 
AM 06 simulation (towards the North).
We illustrate this in Figure 5b, where we show the  
positions of the dark matter clump at 1.34, 1.43, 1.51 and 4.2 Gyr, taken from a merging 4 of the 9 images of Figure 3
of AM06. We overlap part of Figure 5 of Flin \& Krywult
(2006), which illustrates the position of the galaxy sub-clump for a wavelet scale of 129 h$_{75}^{-1}$ kpc.

If we scale the ratio of masses of the main cluster to hat of the DM perturber from the AM06 simulation parameters 
and, conservatively, use for A496 the mass of 4.2$\times$10$^{14}$M$_\odot$ (Durret et al. 2000), the 
perturber should be very massive 
(0.84$\times$10$^{14}$M$_\odot$). This is almost three times more massive than HCG62 (Morita et al. 2006), the brightest
HCG in the Ponman et al. (1996) survey. Such a group, if not unusually depleted of gas,
would easily be detected by current X-ray instruments
at the A496 redshift. 
ROSAT All Sky Survey exposures of that region (R$<$ 50$^\prime$ from A496) fails to detect 
a significant X-ray excess from any extended sources as expected by the gasless dark matter perturber scenario 
described here. The excess count  in a square region 18$^{\prime}$ on the side
centered in Flin \& Krywult's (2006) sub clump is 15$\pm$18 background subtracted counts. 
However RASS exposures are too short ($\sim$250 sec) 
to place any significant constraints on the amount of X-ray emitting gas and future combination of weak 
lensing and deeper X-ray observations of that substructure 
with current satellites should
be able to corroborate this prediction.

\acknowledgments 
We acknowledge support from NASA through {\sl Chandra} award number 
GO 4-5145X, NNG04GH85G and GO5-6139X.  RAD was also partially supported by 
NASA grant NAG 5-3247.  RAD also thanks Yago Ascasibar, Jimmy Irwin, Tatiana F. Lagana, 
Narciso Benitez \& Tracy Clarke
for helpful discussions. 

\begin{deluxetable}{lccccccccc}
\small
\tablewidth{0pt}
\tablecaption{Temperatures and Abundances Across the Cold Arm\tablenotemark{a}}
\tablehead{
\colhead{Region \tablenotemark{b} } &  
\colhead{Temperature}  & 
\colhead{Fe }  & 
\colhead{Si }  & 
\colhead{S } &  
\colhead{ Si/Fe }  &  
\colhead{ S/Fe }  &  
\colhead{ Si/S }  &  
\colhead{$\chi^2$}  \nl
\colhead{} &  
\colhead{(keV)}  & 
\colhead{(solar)}  & 
\colhead{(solar)} &  
\colhead{(solar)} &  
\colhead{(ratio)} &  
\colhead{(ratio)} &  
\colhead{(ratio)} &  
\colhead{d.o.f.} 
}
\startdata
Inner & 3.55$\pm$0.11 & 0.86$\pm$0.08 & 0.88$\pm$0.23 & 1.44$\pm$0.35 & 1.03$\pm$0.29 & 1.68$\pm$0.43 & 0.61$\pm$0.22 & 211/204 \nl
Arm   & 3.08$\pm$0.07 & 0.77$\pm$0.07 & 1.60$\pm$0.20 & 1.43$\pm$0.26 & 2.08$\pm$0.32 & 1.86$\pm$0.38 & 1.12$\pm$0.25 & 176/217 \nl
Outer & 4.68$\pm$0.22 & 0.75$\pm$0.13 & 0.72$\pm$0.57 & 2.16$\pm$0.74 & 0.96$\pm$0.78 & 2.89$\pm$1.10 & 0.29$\pm$0.24 & 167/171 \nl
\tablenotetext{a}{From wabs Vapec model}
\tablenotetext{b}{Shown in Figures 3a,c \& 4a,c}
\enddata
\end{deluxetable}

\begin{figure}
  \plottwo{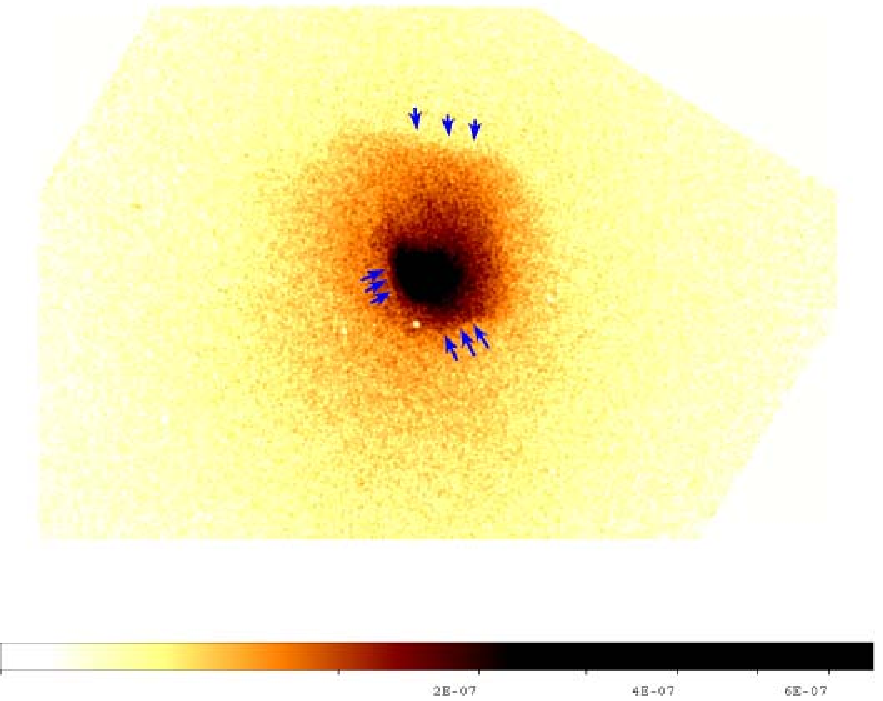}{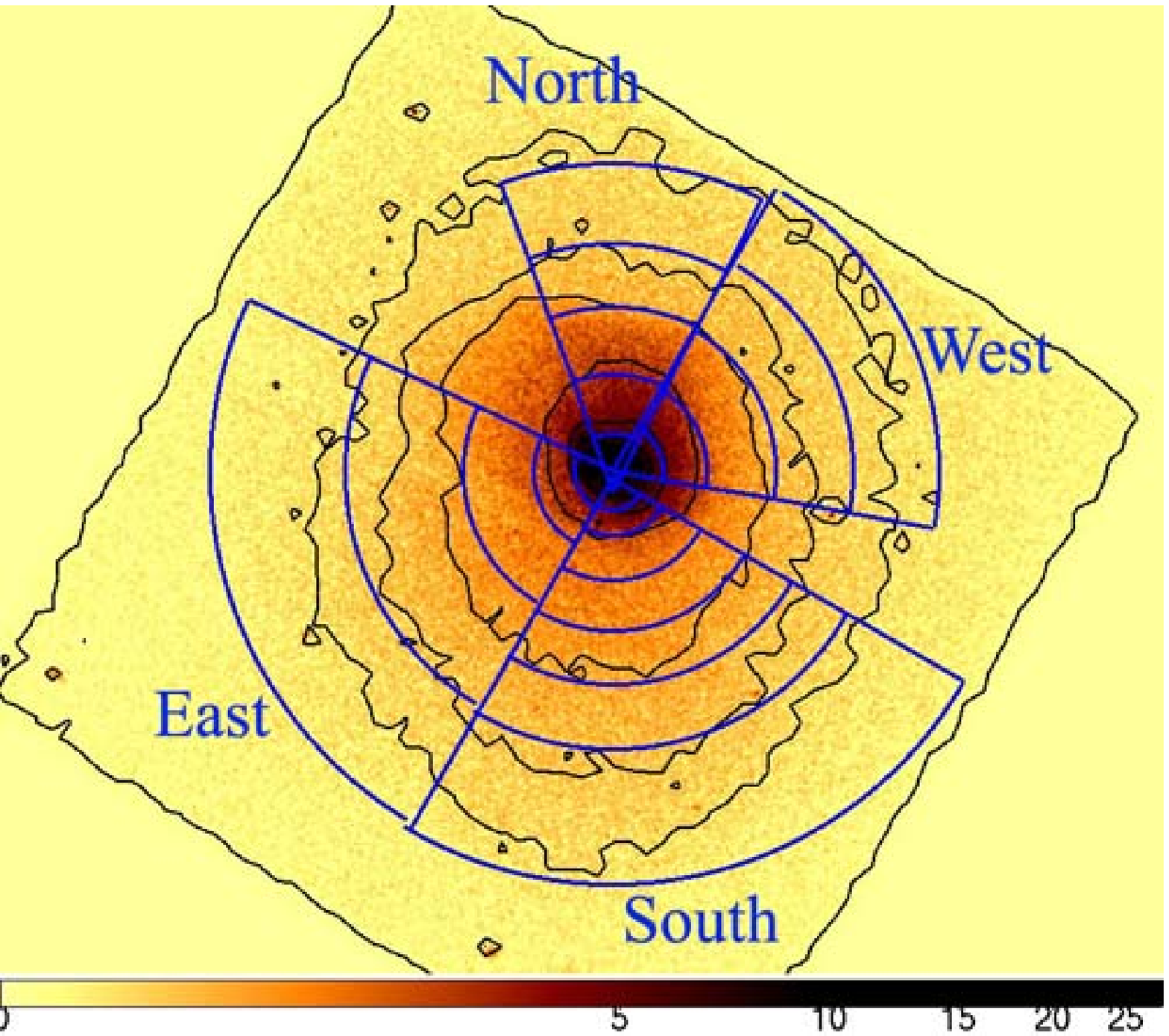}
  \epsscale{0.3}
\caption{(a) Exposure corrected \chandra image of Abell 496 smoothed in DS9 version 4.0 
with a Gaussian function with 3 pixel kernel radius. 
Blue arrows show the position of the northern, southern and 
eastern main cold fronts.  (b) Same as (a0 but with the PIE sectorial extraction regions used in the 
radial analysis of the
cold fronts. X-ray contours are also overlaid and are the same as in Figures 2 and 3.The outermost contour
corresponds to ACIS-S3 chip border.}
\end{figure}

\begin{figure}
  \plottwo{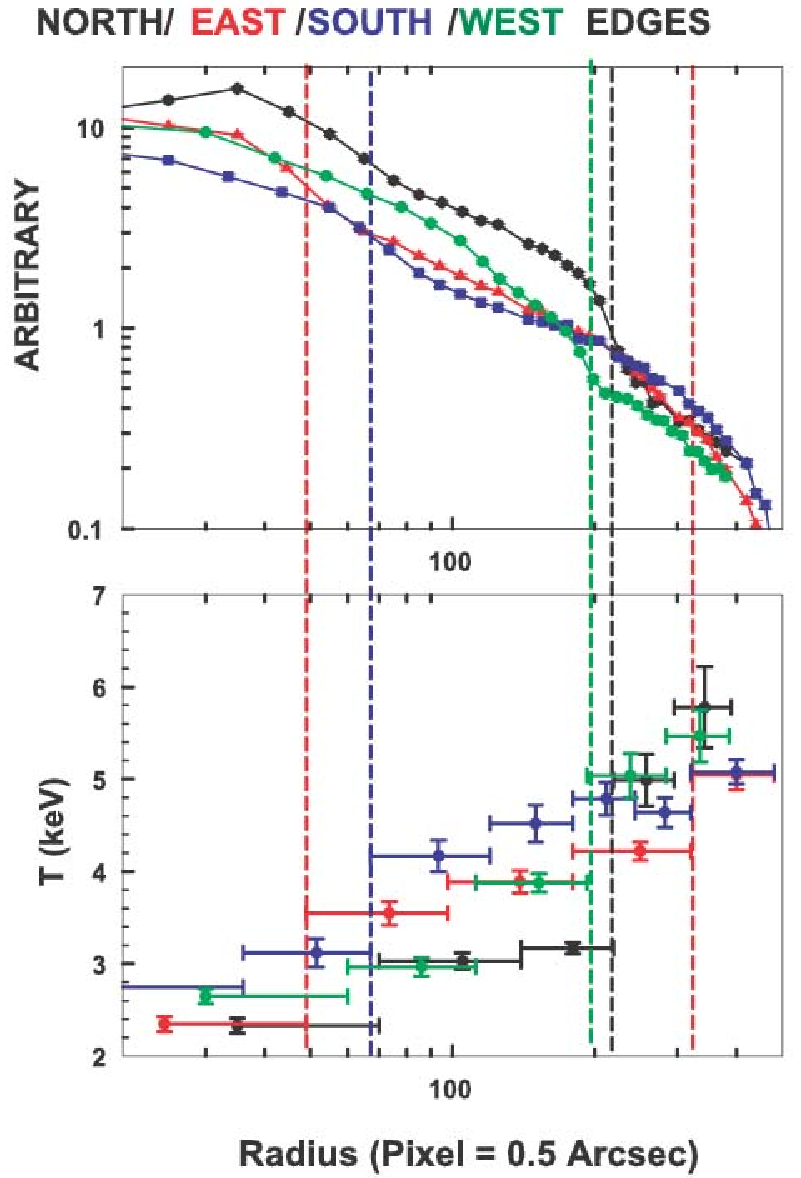}{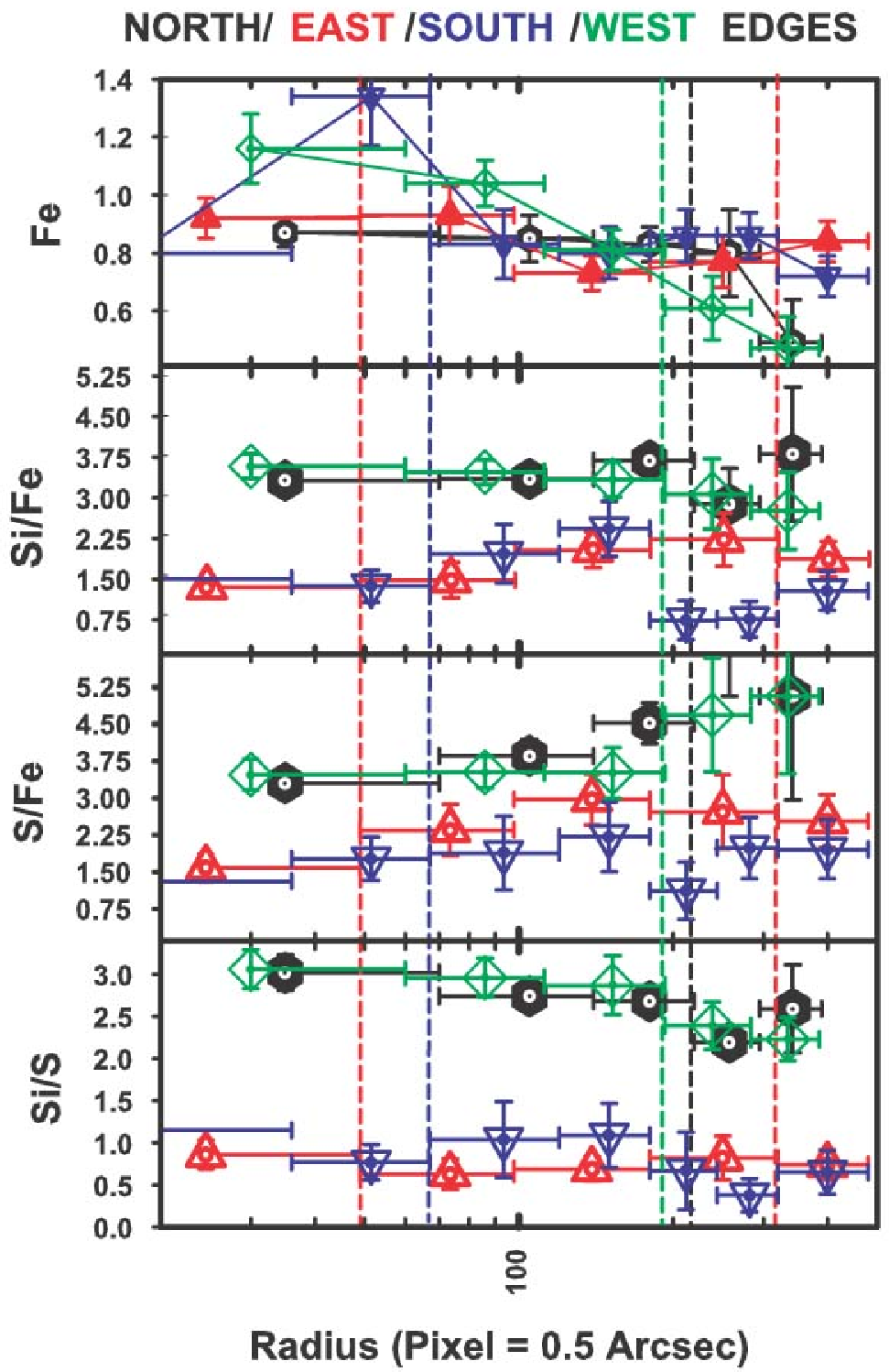}
  \epsscale{0.3}
\caption{(a)Temperature and Density Profiles: Exposure corrected surface brightness {\it Top} and best-fit gas temperature 
{\it Bottom} radial distributions along the 
line of symmetry of the sharp edge (cold fronts) indicated in Figure 1b, 
North (black), East (red), South (blue) and West (green)
 using a {\tt wabs apec} spectral model. The vertical
lines indicate the position of the cold fronts using the same notation for color as the data points.
The extraction regions correspond to those shown in Figure 1b. The units are pixels and 
1 pixel $=$ 0.5$^{\prime\prime} \approx$ 0.33 kpc. Errors
are 1$-\sigma$ confidence. Vertical lines indicate the position of the cold fronts. 
(b) Individual Iron and Metal Abundance Ratio Radial Distributions. Results from spectral fittings using a 
{\tt wabs Vapec} spectral model of same regions as in (a) and the same color code. The values of Si/Fe, S/Fe and 
Si/S for the NORTH (black) and WEST (green) directions were added 2 to their original values for clarity.
The reduced chi-squared for the spectral 
fittings shown are typically 0.7--0.96, with typically  180--250 degrees of freedom.
}
\end{figure}

\begin{figure}
  \plottwo{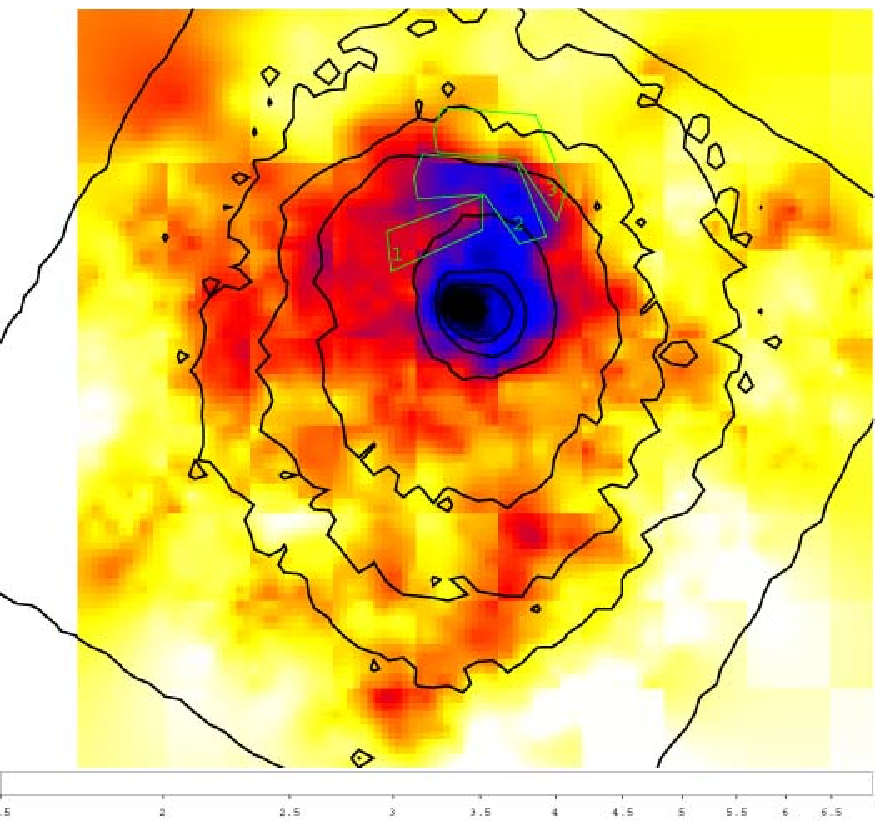}{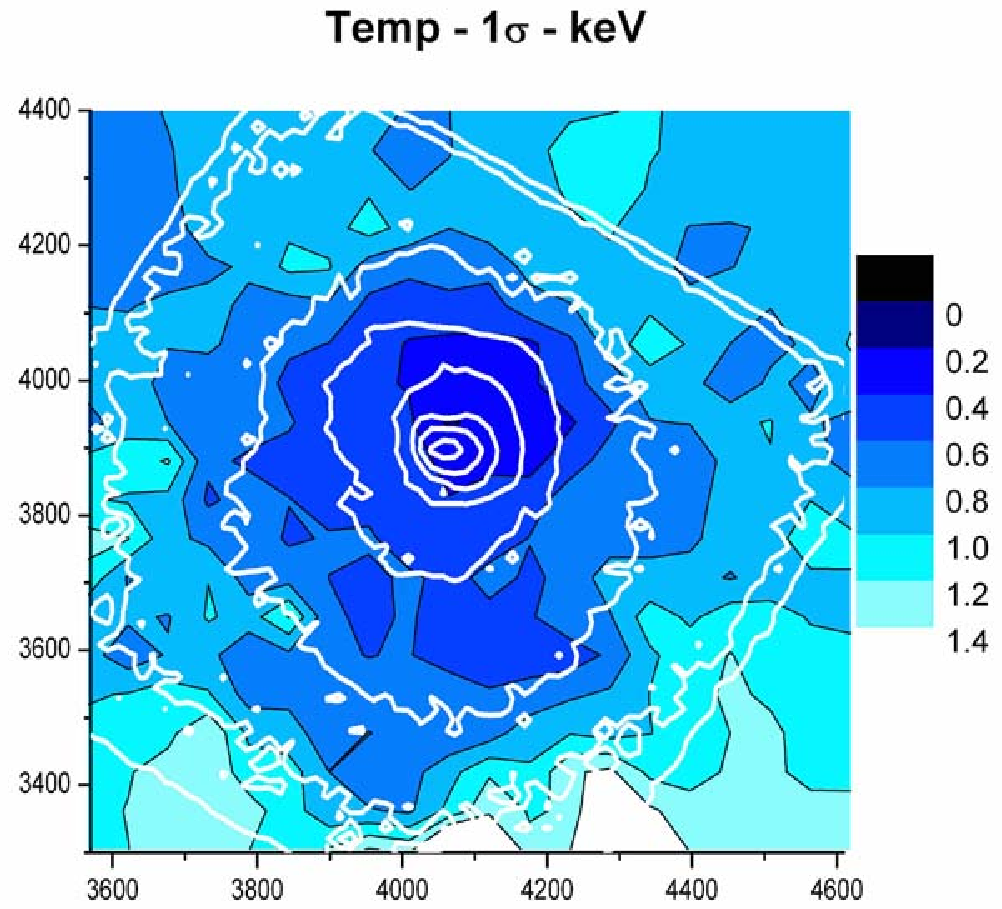}
  \epsscale{0.5}
\end{figure}
\begin{figure}
  \plottwo{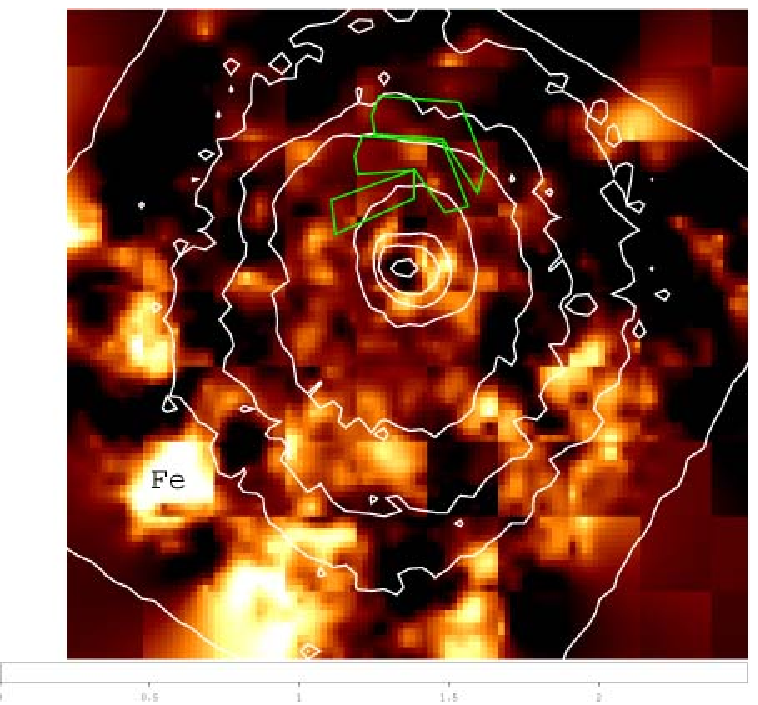}{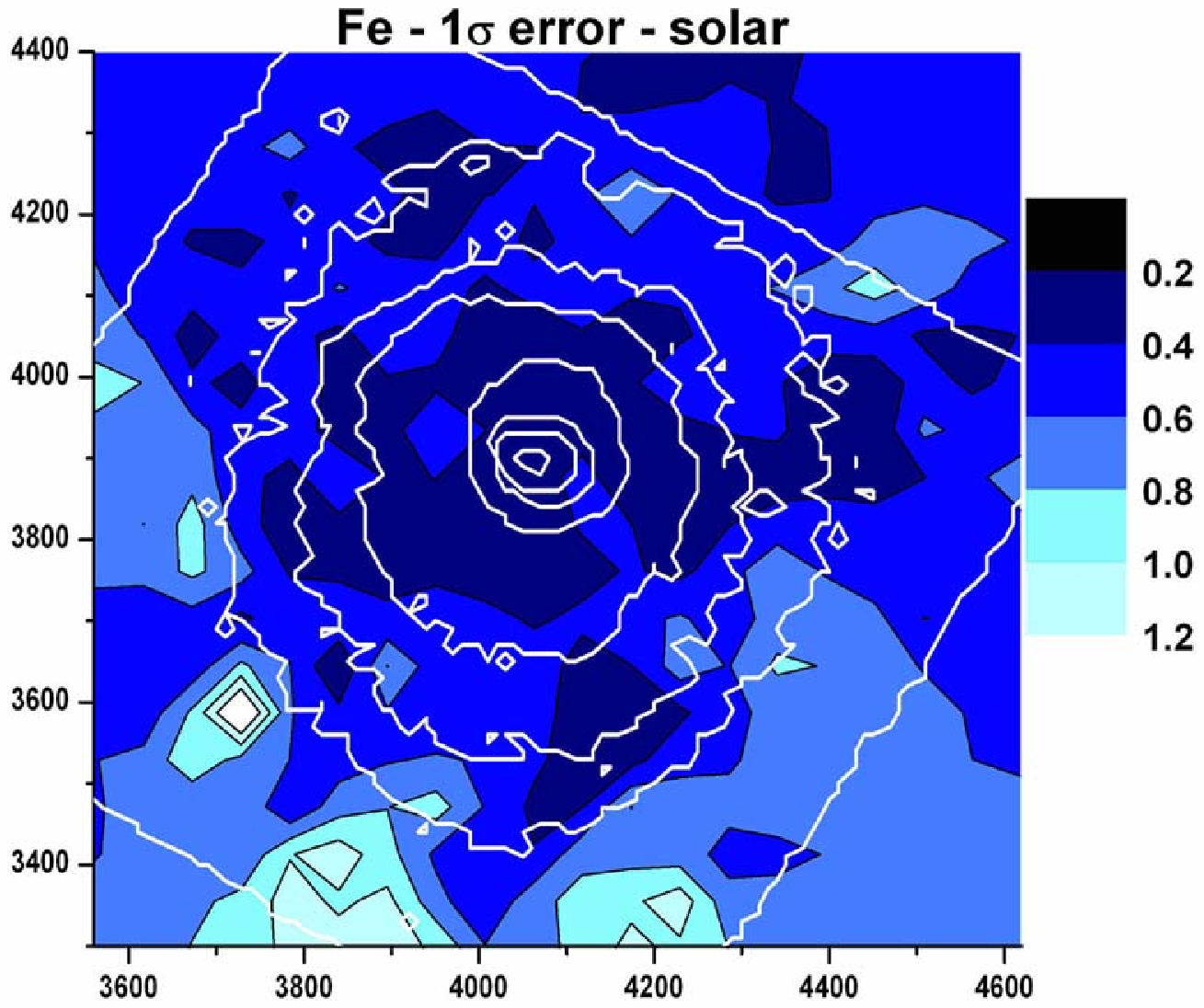}
  \epsscale{0.5}
\end{figure}
\begin{figure}
  \plotone{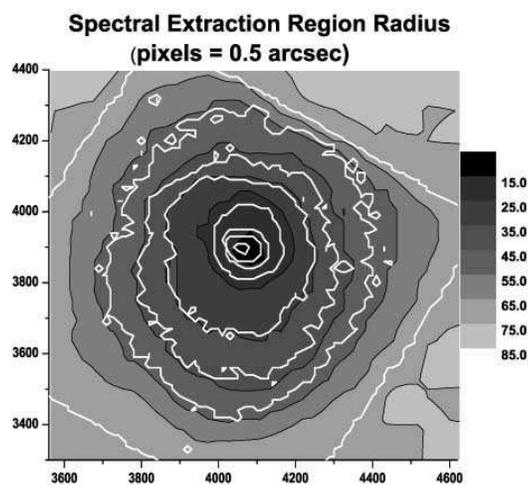}
  \epsscale{0.9}
\caption{Results from an adaptive smoothing algorithm with a minimum of 3000 counts per extraction 
region (circular) and fitted with an absorbed VAPEC spectral model. The gridding method used 
is a correlation method that calculates a new value for each cell in the regular matrix from 
the values of the points in the adjoining cells that are included within the 
search radius, using the Kriging method (e.g. Davis 1986), see section 2 for details.
We also overlay the X-ray contours shown in Figure 1b on top). 
North is up. The units are pixels and 1 pixel $=$ 0.5$^{\prime\prime} \approx$ 0.33 kpc.
The outermost contour corresponds to ACIS-S3 chip border and is centered at RA=68.4084 deg, Dec=-13.261 deg. 
Values outside the CCD borders are an effect of the smoothing algorithm and should be ignored.
The parameters mapped are (a) Temperature \& (c) Abundance. (b) and (d) show the adaptively 
smoothed 1-$\sigma$ error maps for temperatures and abundances, respectively. Figure (e) shows
a color contour plot of the radii of the spectral extraction regions used to determine the parameters shown in 
(a)--(d). It basically gives an idea of the resolution, or smoothing kernel radius, of the 2-D maps above.
                               }
\end{figure}
\clearpage
\begin{figure}
  \plottwo{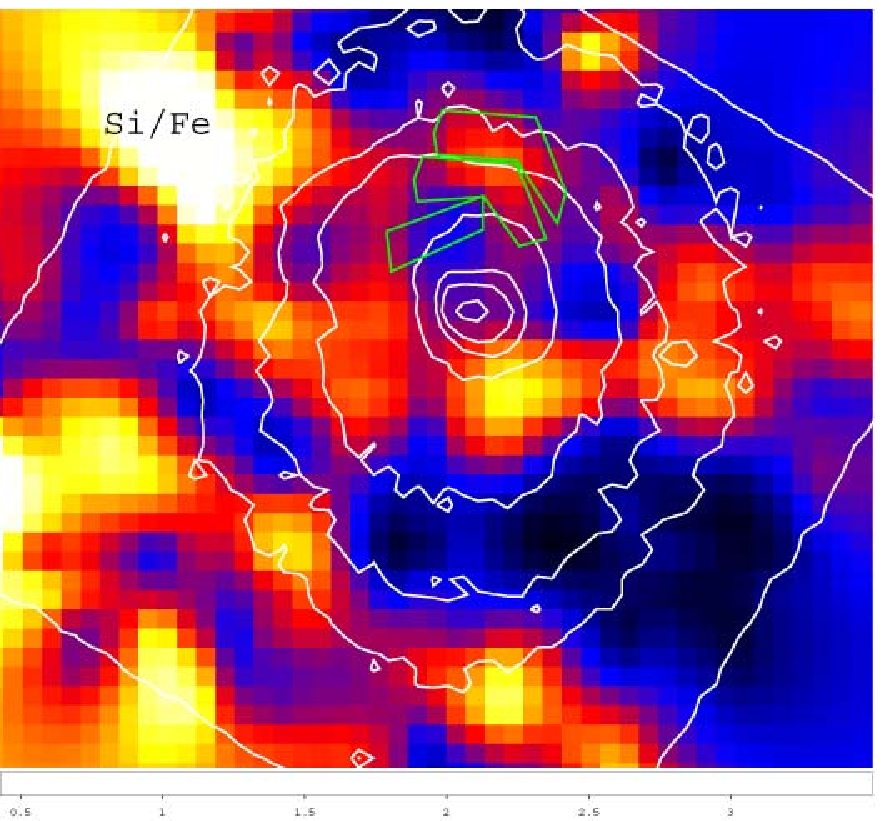}{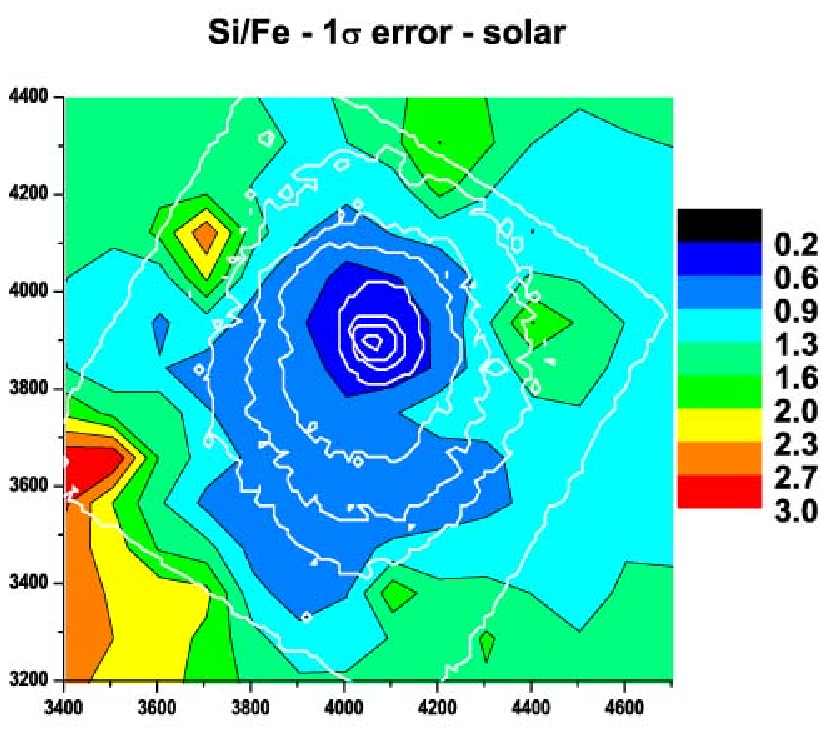}
  \epsscale{0.5}
\end{figure}
\begin{figure}
  \plottwo{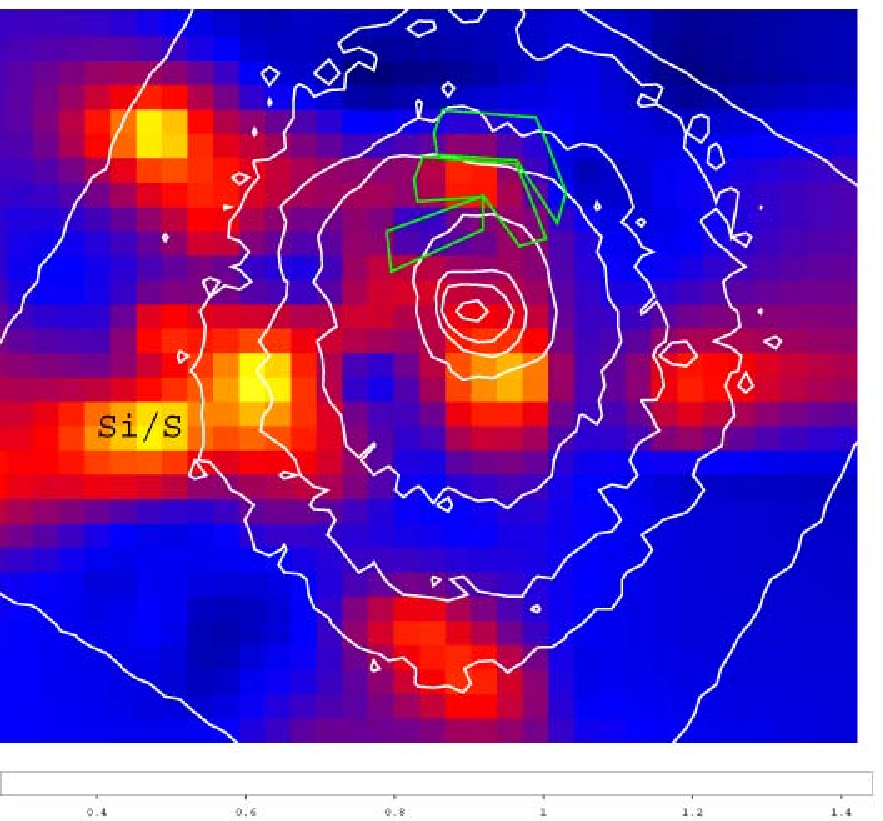}{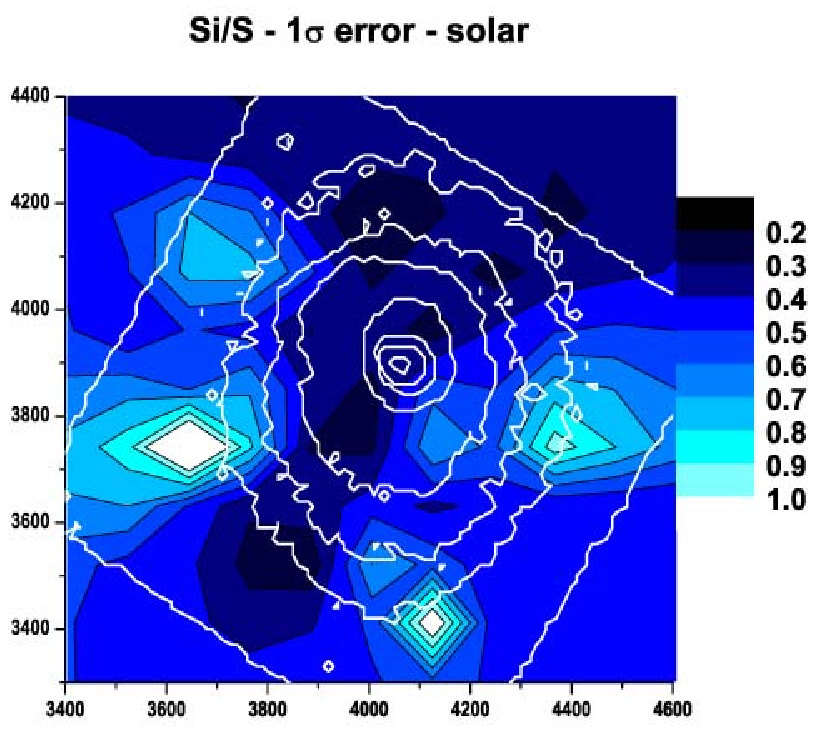}
  \epsscale{0.5}
\end{figure}
\begin{figure}
  \plotone{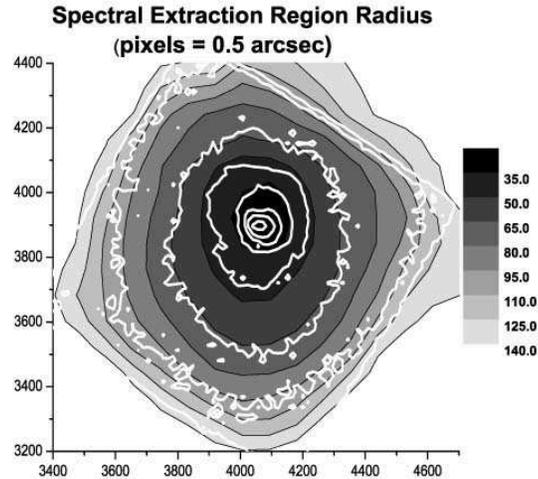}
  \epsscale{0.9}
\caption{Individual Metal Abundance Ratios. Same notation as in Figures 3 but for 
the silicon to iron (a, b) and silicon to 
sulfur (c, d) ratios. (e) shows, analogously, 
the radii of the spectral extraction regions used to determine the parameters in Figures 3(a)--(d) 
				}
\end{figure}

\begin{figure}
  \plottwo{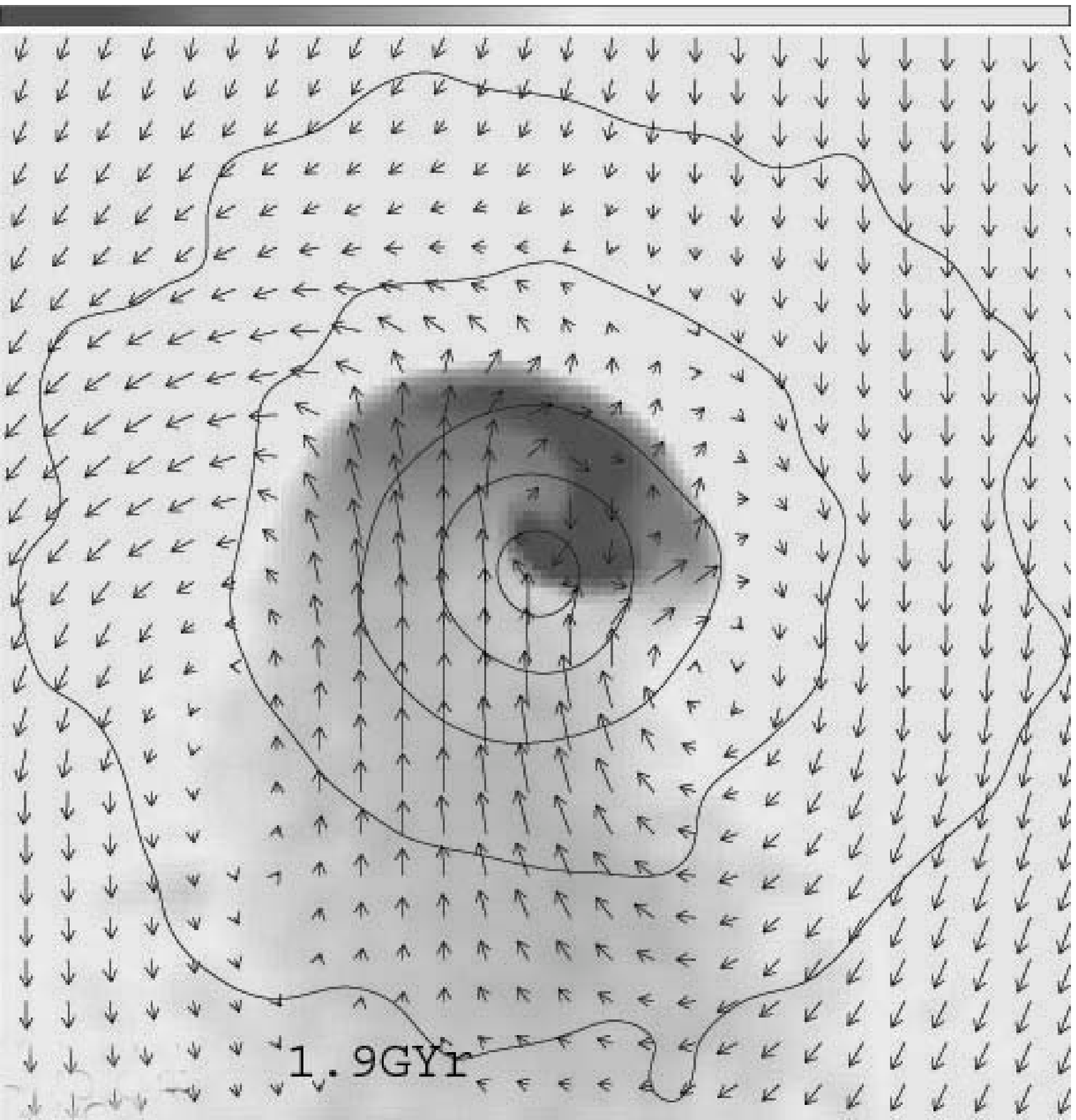}{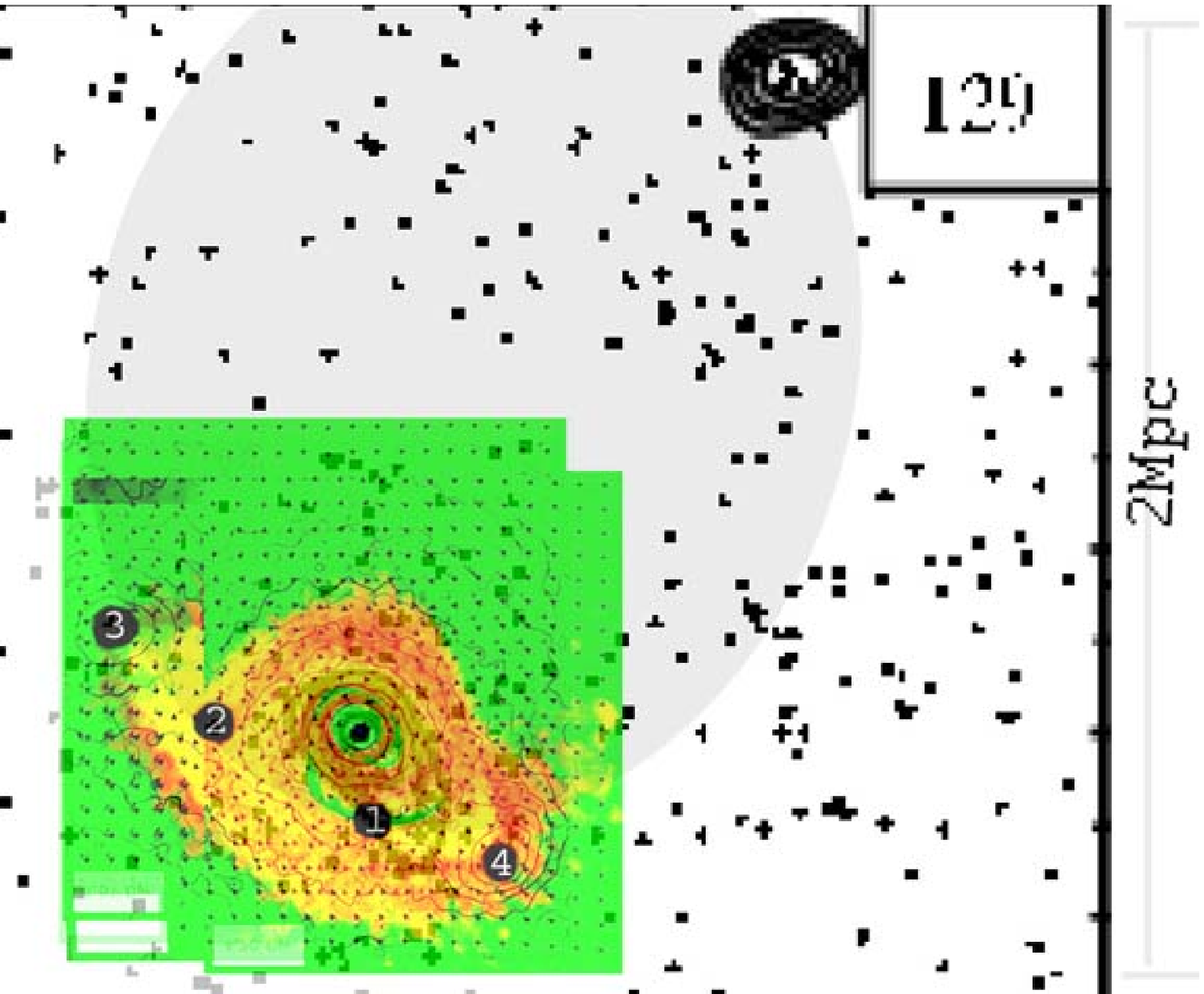}
  \epsscale{0.3}
\caption{(a) zoom in of the core of simulated cluster 0.5 Gyr after the flyby of the dark matter halo. From AM06 (fig.7). Yellow
is $\sim$7--9 keV and blue 2 keV. DM density contours are overlaid and arrows indicate gas velocity (the longest corresponding
to 500 km~s$^{-1}$. The size of the panel is 250 kpc. The figure has been flipped vertically to match the configuration of 
the cold front in A496. (b) Zoom out of the simulation in (a) for epochs 1.34 Gyr, 1.43 Gyr, 1.51 Gyr and 4.2 Gyr, indicated 
by numbers 1,2,3 \& 4, respectively, merged and placed in scale to galaxy distribution map of Flin and Krywult (2006) with 
wavelet image for scale of 129 kpc shown by the contours. 
				}
\end{figure}

\end{document}